\documentclass[a4paper,fleqn]{cas-sc}

\usepackage[utf8]{inputenc}
\usepackage{graphicx}
\usepackage[section]{placeins}
\usepackage{soul}
\usepackage{subcaption}
\usepackage{amsmath}
\usepackage{appendix}
\usepackage[numbers,sort&compress]{natbib}

\def\tsc#1{\csdef{#1}{\textsc{\lowercase{#1}}\xspace}}
\tsc{WGM}
\tsc{QE}
\tsc{EP}
\tsc{PMS}
\tsc{BEC}
\tsc{DE}

\begin{document}
\let\WriteBookmarks\relax
\def\floatpagepagefraction{1}
\def\textpagefraction{.001}

\shorttitle{Accelerating Burgers turbulence simulations with deep learning}

\shortauthors{M. Dhingra et~al.}

\title [mode = title]{Accelerated evolution of Burgers' turbulence with coarse projective integration and deep learning}                      



%
\author[1]{M. Dhingra}[type=editor,
                        orcid=0000-0000-0000-0000]

\cormark[3]


\ead{dmrigank@vt.edu}


\credit{Carrying out the study, Developing deep learning methodology, Data Curation, Writing - Original draft preparation}

\affiliation[1]{organization={Mechanical Engineering, Virginia Tech},
    city={Blacksburg},
    postcode={Virginia 24061}, 
    country={United States}}

\author[2]{O. San}

\affiliation[2]{organization={Mechanical Engineering, Oklahoma State University},
    city={Stillwater},
    state={Oklahoma 74074},
    country={United States}}

\author[3]{A.E. Staples}[%
   ]


\affiliation[3]{organization={Biomedical Engineering and Mechanics, Virginia Tech},
    city={Blacksburg},
    state={Virginia 24061},
    country={Unites States}}


\cortext[cor1]{Corresponding author}



\begin{abstract}
Simulating turbulence to stationarity is a major bottleneck in many engineering problems of practical importance. 
The problem can be cast as a multiscale problem involving energy redistribution processes that take place on the long large eddy turnover time scale and chaotic processes that take the much shorter time scale of the turbulent fluctuations. But the absence of a way to perform super-resolution reconstructions of the instantaneous velocity field from its lower-dimensional moments has prevented the implementation of multiscale computational approaches for accelerating turbulence simulations.
Here, we use an encoder-decoder recurrent neural network to learn the relationship between the instantaneous velocity field and energy spectrum for the one-dimensional stochastic Burgers' equation, a common toy model of turbulence. We use the model to reconstruct the velocity field from instantaneous energy spectra in an equation-free coarse projective integration multiscale simulation scheme using a velocity-energy, fine-coarse multiscale framework. The multiscale-machine learning approach put forth here recovers the final statistically stationary turbulent Burgers' velocity field up to 442 times faster in wall clock time than using direct numerical simulation alone, with 3-digit accuracy.


\end{abstract}


\begin{highlights}
\item A 1D turbulence model is accelerated to stationarity using equation-free multiscale methods
\item Deep learning enables super-resolution
reconstruction of velocity field from energy spectrum data
\item LSTM-based encoder-decoder with attention neural network model used for translation between the two levels of description 
\item Wall clock time savings factors up to 442 are achieved with 3-digit accuracy in the final velocity field
\end{highlights}


\maketitle

\section{Introduction}

High-fidelity turbulence simulations are crucially important, but prohibitively computationally expensive. 
In direct numerical simulations (DNS) of turbulent flows the Navier-Stokes equations are solved without using any turbulence modeling \cite{moinDIRECTNUMERICALSIMULATION1998, abeDirectNumericalSimulation2001}. Solving the Navier-Stokes equations with sufficient temporal and spatial resolution to represent all the scales of turbulence, from the smallest dissipative scales (the Kolmogorov microscales), up to the largest energy-containing integral scales is computationally expensive \cite{jimenezContributionsKolmogorovTheory2004}). The number of floating-point operations required scales as the Reynolds number of the flow raised to the third power \cite{ciofaloDirectNumericalSimulation2022}. For this reason, reduced, averaged, or filtered representations of the Navier-Stokes equations, such as those used in large eddy simulations or in the Reynolds-averaged Navier-Stokes equations, are used almost exclusively to simulate turbulent flows of practical engineering relevance. In these simulation approaches, details of the turbulence are modeled rather than simulated. Turbulence models tend to be heuristic in nature and modeling approaches that are highly accurate for some flows often don't generalize to other flow settings. There is a need for generalized turbulence modeling techniques that do not rely on explicit physics modeling and can be applied in any setting to reduce the computational expense of simulating turbulence.

In recent years, significant advances have been made in the domain of data-driven techniques and machine learning to solve multiscale-based closure problems involving complex dynamical systems. 
Various data-driven approaches have been used to capture the multiscale behavior of these systems. The approaches generally involve using existing experimental or simulation-based data to learn the underlying relationships between the different scales of the system \cite{wangDeepMultiscaleModel2020, alberIntegratingMachineLearning2019, arbabiLinkingMachineLearning2020, yangWhenMachineLearning2020, janssensAdvancingArtificialNeural2022}.

It can be argued that turbulence evolves on two distinct time scales that govern the evolution of the system toward a state of statistical equilibrium. The slow time scale is correlated with the evolution of the shape of the energy spectrum, velocity field structure functions, and other moments of the velocity field to their final equilibrium scalings, and the fast time scale is associated with the rapid fluctuations inherent in a turbulent flow. This multiscale evolution presents an opportunity for the application of multiscale simulation methods that reduce the expense of simulating a system to stationarity. \cite{houMultiscaleModelingIncompressible2013, luoMULTISCALEMODELINGMULTIPHASE2009, engquist2003multiscale, weinan2005nested, weinan2003analysis, fatkullin2006computational}. The DNS of turbulence often applies random forcing to the large scales of motion in the inertial range to produce a statistically stationary velocity field in which the average rate of energy addition to the velocity field via the random forcing is equal to the average energy-dissipation rate at the smallest scales of motion, beginning from (almost) arbitrary initial conditions \cite{eswaranExaminationForcingDirect1988}. Such forcing schemes, when applied to non-stationary velocity initial conditions, immediately produce velocity fields with fast and slow dynamics and associated fast and slow times scales. The fast time scale is associated with the turbulence fluctuations produced by the forcing and is characterized by the Taylor microscale. While the slow time scale is associated with the evolution of the velocity field and its higher-order moments to their statistically stationary forms, and is characterized by the large eddy turnover time. 

Given the multiscale nature of evolving turbulent velocity fields, one should, in principle, be able to apply multiscale computational techniques in order to accelerate the convergence of turbulence simulations to stationarity. But the inability to transform, or super-resolve, coarse variables describing the slow dynamics of turbulence to the fine variables that describe the fast turbulent dynamics, without resorting to explicit and non-generalizable physics modeling, has been a barrier to applying multiscale methods to the problem of turbulence.

Here, we use a machine learning super-resolution model to enable the multiscale simulation of the stochastic Burgers equation (SBE) in one dimension \cite{becBurgersTurbulence2007}. We employ an equation-free coarse projective integration (CPI) multiscale simulation scheme that treats the energy spectrum of the turbulent flow as the coarse, slowly evolving variable and the velocity field as the fine variable that fluctuates on short time scales \cite{kevrekidisEquationFreeMultiscaleComputation2009, zouEquationfreeParticlebasedComputations2006, kevrekidisEquationfreeCoarsegrainedComputational2006}. The CPI scheme reduces the total time taken to evolve the flow to statistical stationarity by taking large, explicit projective time steps in the coarse (energy spectrum) variable, rather than in the fine (velocity field) variable, which would require a smaller time step size for stability and accuracy. Although there is no equation of the form '$dE/dt = f(...)$' that governs the evolution of the energy spectrum for Burgers turbulence that the authors are aware of, the scheme moves between the two levels of description of the system using lifting and restriction operators. The restriction operator transforms instances of the velocity field to lower-dimensional instances of the energy spectrum by integrating the square of the Fourier transform of the velocity field. But going in the other direction, from the coarse level of description to the higher-dimensional fine level of description via a lifting operator, is a difficult task because the energy spectrum does not contain the required velocity field phase information. This task is accomplished in our scheme using a black box deep learning model for the lifting operator that extracts and learns the relational evolution between the two levels of description of the evolving Burgers turbulence. We used a long term-short memory (LSTM) \cite{hochreiterLongShortTermMemory1997} sequence-to-sequence encoder-decoder \cite{sutskeverSequenceSequenceLearning2014} with attention \cite{vaswaniAttentionAllYou2017} mechanism neural network deep learning model.

The paper has been organized in the following manner. In section \ref{eqmm_sec}, we review the existing work on the Equation-free multiscale methods and the coarse projective integration (CPI) scheme. In section \ref{dlbcm}, we discuss how the co-evolving velocity field and energy spectrum descriptions of the flow can be used to accelerate the simulations. 
In subsection \ref{encdec_sec}, we discuss the deep learning architecture and its integration into the CPI scheme. In section \ref{res_sec} and \ref{summ_sec}, we present simulation results and conclude that the deep learning-multiscale simulation approach introduced here significantly reduces the computational expense of evolving the stochastic Burger’s to statistical stationarity while maintaining a high level of accuracy in the velocity field.

\section{Methodology}

Descriptions of the physical model, numerical methods used to solve the model, and computational multiscale and machine learning methods are given in the following sections.

\subsection{Stochastic Burgers' equation}

The Stochastic Burgers' equation (SBE) has been used as a one-dimensional turbulence model. \cite{becBurgersTurbulence2007}. The SBE can be written as
\begin{gather}
    \frac{\partial u}{\partial t} + u\frac{\partial u}{\partial x} = \nu  \frac{\partial^2 u}{\partial x^2} + \eta (x, t) \label{burgers},
\end{gather}

\noindent where $u(x,t)$ is the velocity field, $\nu$ is the viscosity and is taken as $10^{-5}$ in this work, and $\eta (x, t)$ is a noise term (white in time but spatially correlated) of the form
\begin{gather}
    \left \langle \hat{\eta}(k,t)\hat{\eta}({k}',{t}') \right \rangle = 2D_{0}|k|^{\beta }\delta (k + {k}')\delta (t - {t}')
\end{gather}

\noindent where $\hat{\eta}(k,t)$ denotes the spatial Fourier transform of the noise $\eta (x, t)$ and $D_{0}$ and $\beta$ are the amplitude and spectral slope of the noise. For the SBE to portray complex multifractal behavior characteristic of turbulence, the $\beta$ value must lie between $-1$ and $-2/3$  \cite{basuCanDynamicEddyviscosity2009}. The values for $D_{0}$ and $\beta$ for this study have been selected as $10^{-6}$ and $-3/4$. The noise term is
\begin{gather}
    \eta(x,t) = \sqrt{2D_{0}/\Delta t}{\mathcal{F}}^{-1}(|k|^{\beta/2}\hat{f}(k)),
\end{gather}

\noindent where $\hat{f}(k)$ is the Fourier transform of a Gaussian random variable $f(x)$ with mean and standard deviation of $0$ and $\sqrt{N}$. The notation ${\mathcal{F}}^{-1}$ denotes the inverse Fourier transform; $N$ is the number of grid points ($2^{12}$ modes for this study). The direct numerical simulations (DNS) are run for 2 million time steps with an Euler explicit scheme for the first time step and 2nd order Adams-Bashforth for the rest of the time steps with $\Delta t = {10}^{-4}$. The domain length $L$ has been taken as $2\pi$ where the nodal spacing $\Delta x$ is $2\pi/N$. For this study, we assign the time taken to reach 1 million time steps as the total time required for the system to reach a statistically stable state, which is approximately equivalent to the eddy turnover time \cite{basuCanDynamicEddyviscosity2009}. Eddy turnover time provides a measure of how long it takes for energy to cascade through the turbulent flow from large to small scales.

\subsection{Equation-free multiscale methods} \label{eqmm_sec}

Equation-free (EF) modeling is a data-driven simulation technique that uses computational data from a different level of description of the system to approximate the behavior of a complex system on a level of description of interest, rather than relying on a set of equations that describe the system's dynamics on the level of interest \cite{kevrekidisEquationFreeMultiscaleComputation2009, zouEquationfreeParticlebasedComputations2006, kevrekidisEquationfreeCoarsegrainedComputational2006, junyichengMultiscaleMethodEquation2018, cazeauxProjectiveMultiscaleTimeintegration2019, alexanderEquationfreeImplementationStatistical2008}. This approach is particularly useful when the underlying evolution equation of a system in the variable of interest is unknown or is too complex to be solved using analytical or computational techniques. EF modeling generally involves collecting data from simulations (or experiments) and using it to model an approximation function that can predict the system's behavior. In our simulations, we have assumed the Burgers' velocity flow field information as the fine-scale level of description of the system and the energy spectrum information as the coarse-scale level of description of the system.

\subsubsection{Coarse projective integration}

For a complex system that evolves on two disparate coarse and fine time scales  \cite{houMultiscaleModelingIncompressible2013, luoMULTISCALEMODELINGMULTIPHASE2009, engquist2003multiscale, weinan2005nested, weinan2003analysis, fatkullin2006computational}, the fine or microscopic scale generally has a relatively fast evolution cycle and is referred to a high dimensional quantity. The coarse or macroscopic scale has a slower evolution cycle and generally exists as a lower dimensional quantity. In order for us to enable EF modeling for a given system, a coarse projective integration (CPI) scheme must be devised \cite{kevrekidisEquationFreeMultiscaleComputation2009, zouEquationfreeParticlebasedComputations2006, kevrekidisEquationfreeCoarsegrainedComputational2006}. The CPI scheme is mainly responsible for transitioning between the two scales of the system until a statistically steady state is achieved. A CPI scheme consists of the following steps:

\begin{enumerate}
    \item Short calls to a microscopic flow simulator.
    \item Restricting the fine-scale information to a coarser scale with a restriction operator ($\mathcal{M}$).
    \item Lifting the projected coarse scale information back to the microscale with a lifting operator ($\mu$).
    \item The lifted information is healed back onto the manifold.
    \item Steps 1-5 are repeated until a statistically stable state at $t_{final}$ is achieved. This state is attained when the turbulent kinetic energy in the system does not fluctuate drastically.
\end{enumerate}

\subsubsection{The restriction operator}

Once the fine-scale velocity information is obtained via DNS, we can make use of a restriction operator $(\mathcal{M})$ to obtain the coarse-scale energy spectrum (or the 2nd-order structure function) \textbf{(Fig. \ref{fig:CPI})} from the velocity data. The restriction operator ($\mathcal{M}$), restricts the microscopic scale information at $u(x, t_{n})$ and $u(x, t_{n+1})$ to $E_{N}$ and $E_{N+1}$.
The restriction operator we use $\mathcal{M}$, is defined as
\begin{gather}
    E_{n}(k) = \mathcal{M} u(x, t_{n});  n = 1, 2, 3,..,N \\
    \mathcal{M}u(x, t_{n}) = \frac{1}{2}\left(\mathcal{F}(\overline{u_{n}})\right)^{2},
\end{gather}

\noindent where $\mathcal{F}$ is the fast Fourier transform of the velocity signal \cite{fourierTheorieAnalytiqueChaleur2009, CooleyTukeyFastFourier2017} and $n$ represents the time step of the velocity signal.

\subsubsection{Coarse scale projection} \label{csp}

We then use an Euler explicit projection function to estimate the coarse scale information at a future time step \textbf{(Fig. \ref{fig:CPI})} $\tilde{E}_{N+2}(k)$ \cite{eulerMethodusInveniendiLineas1744, brenanOneStepMethods1995}. The Euler explicit projective function is defined as
\begin{gather}
    \tilde{E}_{N+2}(k) = E_{N+1} + \Delta T \left(\frac{E_{N+1} - E_{N}}{t_{N+1} - t_{N}}\right) \\
    \left(\frac{\Delta T}{t_{N+1} - t_{N}}\right)  \geq{1}, 
\end{gather}

\noindent where $\Delta T$ is the projective step size associated with the coarse scale; $\Delta T / \delta t$ has to be greater than or equal to 1 in order for the coarse scale projection to result in savings in computational time. 

\subsubsection{The lifting operator}

The coarse-scale projection function in subsubsection \ref{csp} provides an approximation $\tilde{E}(k)$ to the energy spectrum at a future time after a projective step of size $\Delta T$. Using the lifting operator ($\mu$), we can convert the macroscopic scale information contained in $\tilde{E}_{N+2}(k)$ to an intermediate microscopic scale velocity signal \textbf{(Fig. \ref{fig:CPI})} $\tilde{u}(x, t_{N+2})$. The lifting operator ($\mu$) is given as
\begin{gather}
        \tilde{u}(x, t_{N+2}) = \mu \left (\tilde{E}_{N+2}(k)\right),
\end{gather}

\noindent where $\tilde{E}_{N+2}(k)$ is the projected energy spectrum and $u(x, t_{N+1})$ is the velocity field available at the previous time step. In our study, the lifting operator ($\mu$) is assigned to a data-driven encoder-decoder LSTM-based seq2seq model with attention which encodes the coarse scale information at a given time step, generates a context based on the encoding and decodes on the fine-scale information with the help of the generated context \cite{hochreiterLongShortTermMemory1997, sutskeverSequenceSequenceLearning2014, vaswaniAttentionAllYou2017}. The lifting operator is described in subsection \ref{dlbcm}. 

\subsubsection{Healing the lifted information}

\begin{figure}[!h]
    \centering
    \includegraphics[width=1\textwidth]{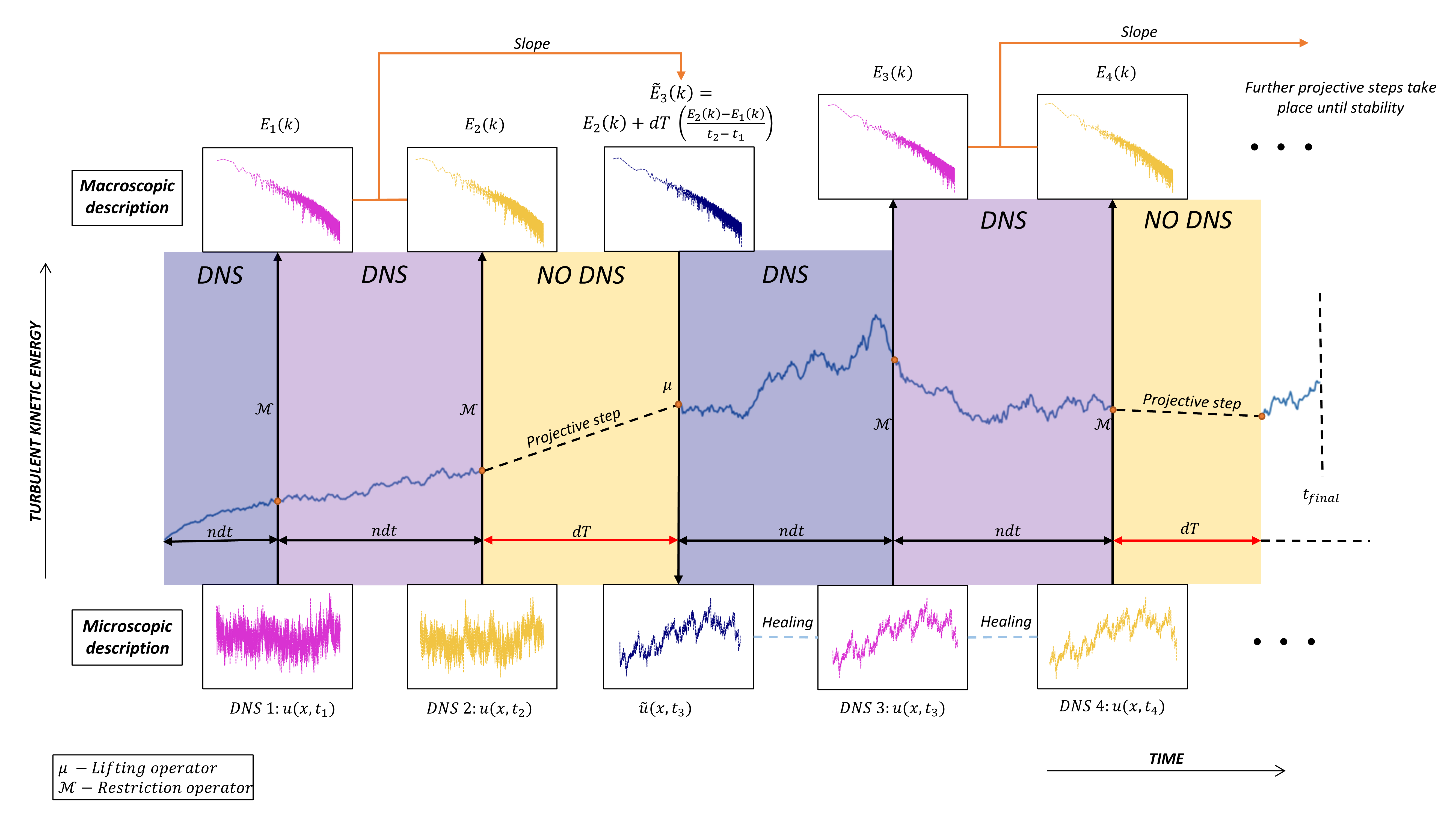}
    \caption{Total kinetic energy versus time for two complete coarse projective integration (CPI) cycles. Lavender and purple boxes depict bursts of DNS, yellow boxes depict coarse projective steps. A complete CPI cycle consists of a lavender, a purple, and a yellow box. In each CPI cycle, the two final DNS velocity fields from the lavender and purple DNS bursts are restricted to find the (coarse-scale) energy spectra at these times; the energy spectra are then used to approximate $dE(k,t)/dt$ and project $E(k,t)$ forward in time by the macroscopic time step, $dT$; the projected spectrum is then lifted to its corresponding velocity field, ending the CPI cycle. Errors in the projected velocity field are healed in the DNS bursts undertaken at the beginning of the next CPI cycle. CPI cycles are undertaken until the final simulation time $t_{final}$ is reached and the flow has achieved a statistically steady state.}
    \label{fig:CPI}
\end{figure}

Errors in the lifted velocity field are ``healed'' by evolving it with the microscopic flow simulator (Eq. \ref{burgers}) until the lifted field approaches the \textbf{(Fig. \ref{fig:CPI})} original solution manifold. The healing step of the CPI cycle consists of initializing the microscopic flow simulator with $\tilde{u}(x, t_{N+2})$ and running the simulation for $n_{heal}$ time steps to a final healing time of $t_{heal}$.

\FloatBarrier

\subsection{Deep Learning-based lifting operator} \label{dlbcm}

In this study, for multiscale-based closure modeling, we employ a commonly used architecture in the field of natural language processing (NLP) known as a sequence-to-sequence (seq2seq) learning model as our lifting operator $\mu$ \cite{sutskeverSequenceSequenceLearning2014, vaswaniAttentionAllYou2017, devlinBERTPretrainingDeep2019, brownLanguageModelsAre2020, bahdanauNeuralMachineTranslation2016}. This architecture is often used for time-series data forecasting, image captioning, language translation, and similar tasks. The model consists of encoder and decoder blocks which are primarily made up of long short-term memory (LSTM) cells placed in a sequence. It also has an 'attention' mechanism which helps the decoder focus on different parts of the input sequence at each step of the decoding process. When an input is given to the encoder block, a context vector or an encoded representation of the model is generated which is then fed to a decoder block to generate the desired output. One of the biggest advantages of a seq2seq model is that the input and desired output vectors need not be of the same or fixed size. A brief description of the seq2seq neural network is as follows

\begin{figure}[h!]
    \centering
    \includegraphics[width=1\textwidth]{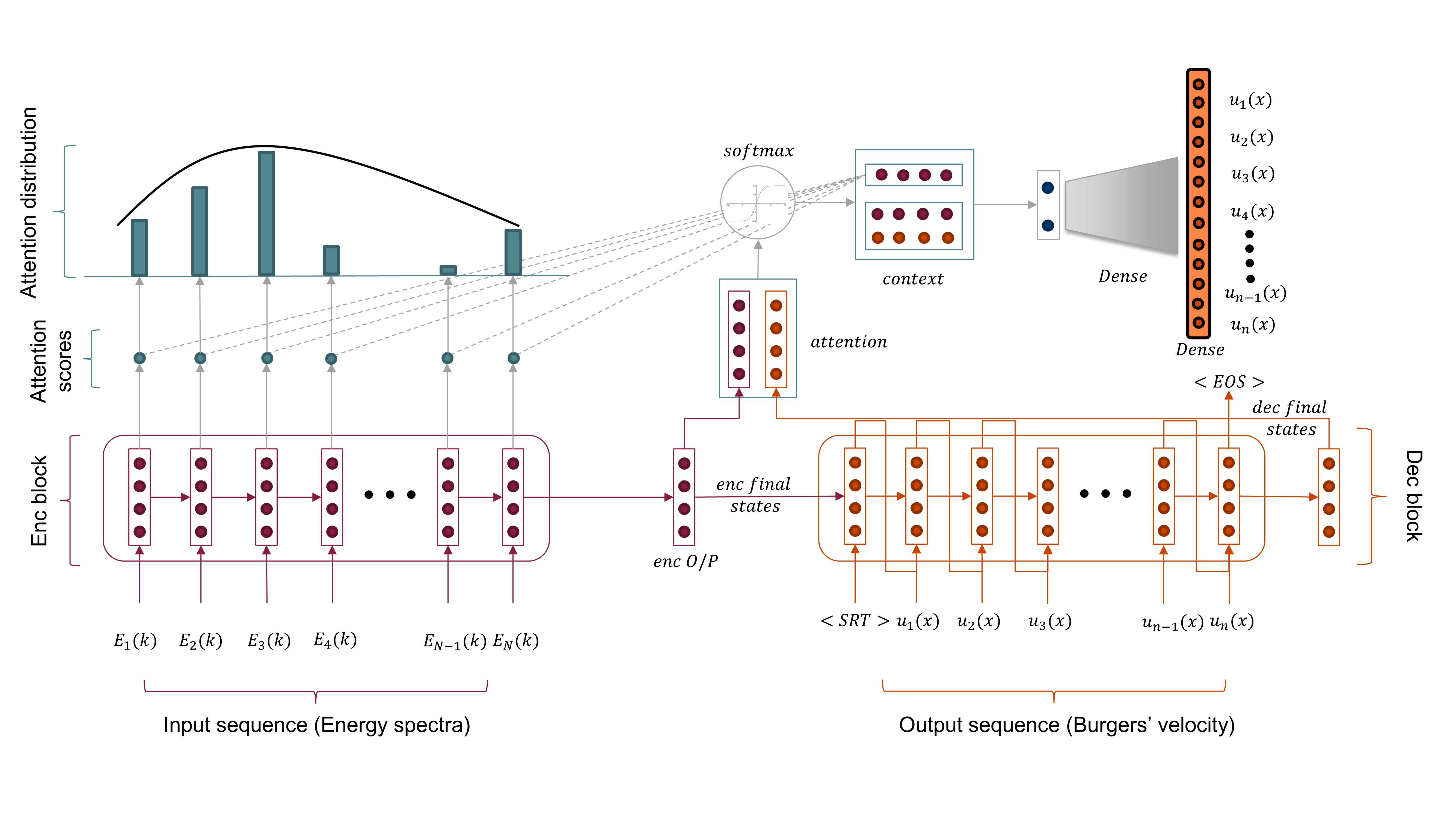}
    \caption{The attention mechanism architecture. The attention layer takes the output from the encoder and decoder at a given instance and calculates the context between the input sequence and the output from the decoder. The context is then funneled through dense layers toward the target output sequence.}
    \label{fig:attn}
\end{figure}

Let $x_{1}, x_{2}, x_{3},..., x_{T-1}, x_{T}$ be an input sequence $X$ where $x_{i}$ is the $i^{th}$ element and $T$ is the length of the given input sequence. The encoder, to which the input sequence is fed, generates a hidden state using the equation
\begin{gather}
    H_{t}(encoder) = \phi(W_{HH}H_{T-1}+W_{HX}H_{T})
\end{gather}

\noindent where $\phi$ is a non-linearity or activation function, $H_{t}(encoder)$ is the hidden states in the encoder, $W_{HH}$ is the weight matrix connecting the hidden states, and $W_{HX}$ is the weight matrix which connects the input to the hidden states.
The output generated by the encoder is given by
\begin{gather}
    H = (h_1, h_2, ..., h_T), 
\end{gather}
\noindent where H denotes the hidden state output for each time step. In the presence of the attention mechanism \textbf{Fig. \ref{fig:attn}}, the decoder computes the current hidden state $s_{t}$ and the current output $Y_t$ as follows
\begin{gather}
    context\_ vector = attention(H, s_{t-1}(decoder)).   
\end{gather}
The hidden states in the decoder are processed using the equation
\begin{gather}
    H_{t}(decoder) = \phi(W_{HH}H_{T-1}(decoder) + W_{CH}(context \_ vector)).
\end{gather}
Here, $context$\textunderscore$vector$ is computed by the attention mechanism. $W_{CH}$ is the weight matrix that connects the context vector to the hidden states in the decoder. The output generated by the decoder is given by
\begin{gather}
    Y_{t} = H_{t}(decoder)W_{HY},  
\end{gather}
\noindent where $W_{HY}$ is the matrix which connects the hidden states with the decoder output. The attention mechanism computes a context vector $ct$ that is used to weigh the importance of each hidden state in the encoder. The context vector is computed as follows
\begin{gather}
    e_{t} = f(s_{t-1}(decoder), h_{t}(encoder)) \\
    a_{t} = softmax(e_{t}) \\
    contex\_vector = Dot(a_{t}H).
\end{gather}
Here, $e_{t}$ is the attention score which captures the similarity between the current decoder hidden state and each encoder hidden state, $f$ is a function that computes the attention score, $softmax$ is the activation function used to normalize the attention scores into a probability distribution, and $Dot$ is a dot product of the encoder hidden state with the attention weights $a_t$. This generates a context vector that provides a summary of relevant information from the input sequence, taking into account the current state of the decoder.

\subsubsection{Data preparation and pre-processing}

For the seq2seq model to understand and learn the important features between the energy spectrum and velocity field information, it is essential to prepare, pre-process, and reshape the data according to the model's input and output requirements.

\begin{enumerate}
    \item Burgers' data generation and collection: In order for us to train a model, we would need to collect a sufficient amount of training data for both, the 1-dimensional Burgers' velocity at different time steps and their corresponding energy spectrum data points. 
    \item Pre-processing and reshaping: We then standardize both datasets by subtracting the standard deviation from and dividing the mean of the entire signal with its data points. The standardization is performed at every time step individually to preserve the underlying dynamics and the flow properties that change over the evolution of the flow field. This also prevents any data leakage from the train dataset into the test dataset as the statistical properties in the train set do not influence the scaling in another set. Let $E_{i,k}$ be the energy spectrum value for time step $i$ and wavenumber $k$. The standardized energy spectrum value $E'_{i,k}$ can be computed as
    
    \begin{gather}
        E'_{i,k} = \frac{E_{i,k} - \mu_{E_i}}{\sigma_{E_i}} \label{E_scale},
    \end{gather}
    \noindent where $\mu_{E_i}$ is the mean of the energy spectrum at time step $i$, and $\sigma_{E_i}$ is the standard deviation of the energy spectrum at time step $i$.
    
    Let $V_{i,j}$ be the velocity field value for time step $i$ and spatial index $j$. The standardized velocity field value $V'_{i,j}$ can be computed as
    \begin{gather}
        V'_{i,j} = \frac{V_{i,j} - \mu_{V_i}}{\sigma_{V_i}} \label{v_scale},
    \end{gather} 
    \noindent where $\mu_{V_i}$ is the mean of the velocity field at time step $i$ and $\sigma_{V_i}$ is the standard deviation of the velocity field at time step $i$.
    For an LSTM-based model, the input shape should be \textbf{(number \textunderscore of \textunderscore samples, time steps, input \textunderscore features)}.
\end{enumerate}

\subsubsection{Model training}

The input to the encoder is the pre-processed energy spectrum information and the input to the decoder is the corresponding velocity field information with ``start of the sequence'' and ``end of the sequence'' labels (as seen in Fig. \ref{fig:attn}). The ``TimeDistributed'' dense output layer is assigned to the same velocity information at that given time step. This method is commonly known as ``teacher forcing'' in the domain of seq2seq learning. The training is carried out until an acceptable validation loss is reached via an early stopping function call. We also incorporate a learning rate scheduler that reduces the learning rate by a factor of 10 when the evaluation loss becomes stagnant. This ensures a stable learning cycle.

\subsubsection{Model inference and data post-processing}

The trained seq2seq model expects the projected energy spectrum signal at the encoder input. The velocity field information is reconstructed at the decoder output. The scaled reconstruction can be scaled back to its original values with the help of a statistical or linear regression model that can learn the scaling values from the equations \ref{E_scale} and \ref{v_scale}. Let $E'_i$ be the standardized projected energy signal at time step $i$, and let $\mu_{E_i}$ and $\sigma_{E_i}$ be the mean and standard deviation of the energy signal at time step $i$. We can use a Random Forest Regressor model $F$ to predict the mean $\mu_{V_i}$ and standard deviation $\sigma_{V_i}$ of the velocity field at time step $i$:

\begin{gather}
    (\mu_{V_i}, \sigma_{V_i}) = F(E'_i, \mu_{E_i}, \sigma_{E_i})
\end{gather}
Unscaled signal can then be calculated as follows
\begin{gather}
    V_{i_{lift},j} =  \sigma_{V_i}V'_{i_{lift},j} + \mu_{V_i},
\end{gather}
\noindent where $V'_{i_{lift},j}$ is the scaled lifted velocity signal and $V_{i_{lift},j}$ is the unscaled lifted signal.

\subsubsection{The encoder-decoder model} \label{encdec_sec}

In this study, we employ a deep learning model based on an encoder-decoder architecture with an attention mechanism, as shown in Fig. 3. The details of the model components are as follows.

\begin{enumerate}
    \item Encoder: The encoder consists of single or multiple layers of LSTM cells with 128 hidden units and a rectified linear unit (ReLU) activation function. Dropout and recurrent dropout rates are set to 0.2 to prevent overfitting. Both return \textunderscore sequences and return \textunderscore state parameters are set to ``True'' to obtain all hidden states and the last hidden state of the encoder, respectively.
    \item Batch Normalization: After the encoder, a batch normalization layer is added with a momentum value of 0.5. This layer reduces internal covariate shift, improves stability during training, speeds up convergence, and enhances model accuracy. The batch normalization process is represented as: 
    \begin{gather}
        y = \gamma(x-\overline{x})/\sigma + \omega,    
    \end{gather}
    where $x$ is the activation of a layer, $\overline{x}$ and $\sigma$ are the mean and standard deviation of the activations, $\gamma$ and $\omega$ are learnable scale and shift parameters, and $y$ is the output after normalization.
    \item Decoder: The decoder is initialized with the decoder inputs and the hidden states of the encoder outputs. It also employs a ReLU activation function. The batch normalized decoder outputs and encoder outputs are passed through the attention mechanism, followed by a concatenate layer.
    \item TimeDistributed Layer: The output of the attention cell is passed through another BatchNormalization layer and a TimeDistributed layer with a single dense layer, which applies a dense layer to each time step of the input tensor.
\end{enumerate}
The model is compiled using the Adam optimizer with a learning rate of $7e-4$ and a gradient clipping norm of 1. The loss function used is a custom function that penalizes the model output on the sum of mean squared error (MSE) and mean absolute error (MAE).

\begin{gather}
    \text{Custom Loss} = \text{MSE} + \text{MAE}
\end{gather}
\noindent where
\begin{gather}
    \text{MSE} = \frac{1}{n} \sum_{i=1}^{n} (y_i - \hat{y}_i)^2 \\
    \text{MAE} = \frac{1}{n} \sum_{i=1}^{n} |y_i - \hat{y}_i|
\end{gather}

\begin{figure}[!h]
    \centering
    \includegraphics[width=1\textwidth]{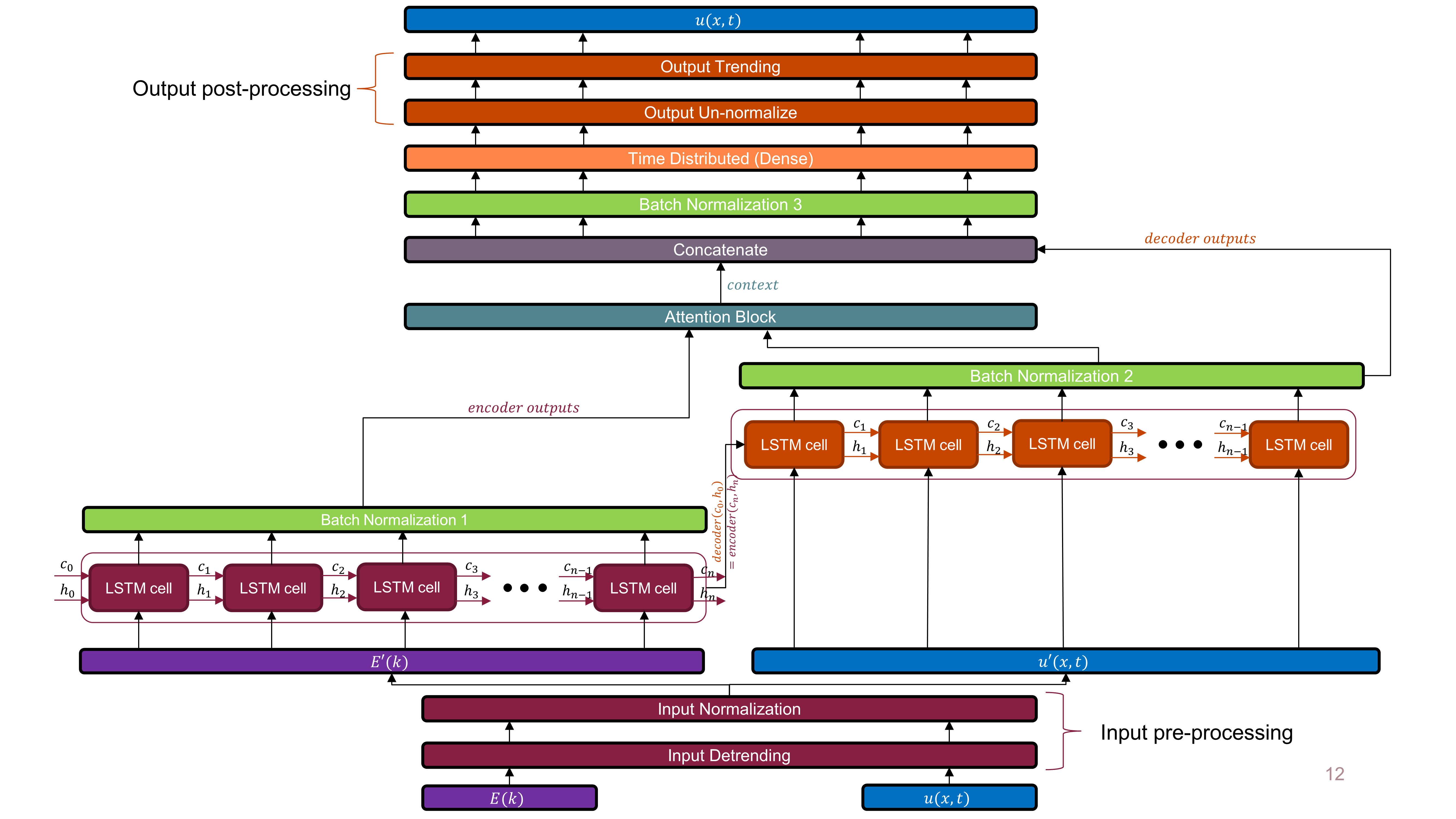}
    \caption{The architecture of the LSTM-based encoder-decoder model with attention.}
    \label{fig:encdec}
\end{figure}

\section{Results and Findings} \label{res_sec}
    
\begin{figure}[!h]
        \centering
        \includegraphics[width=1\textwidth]{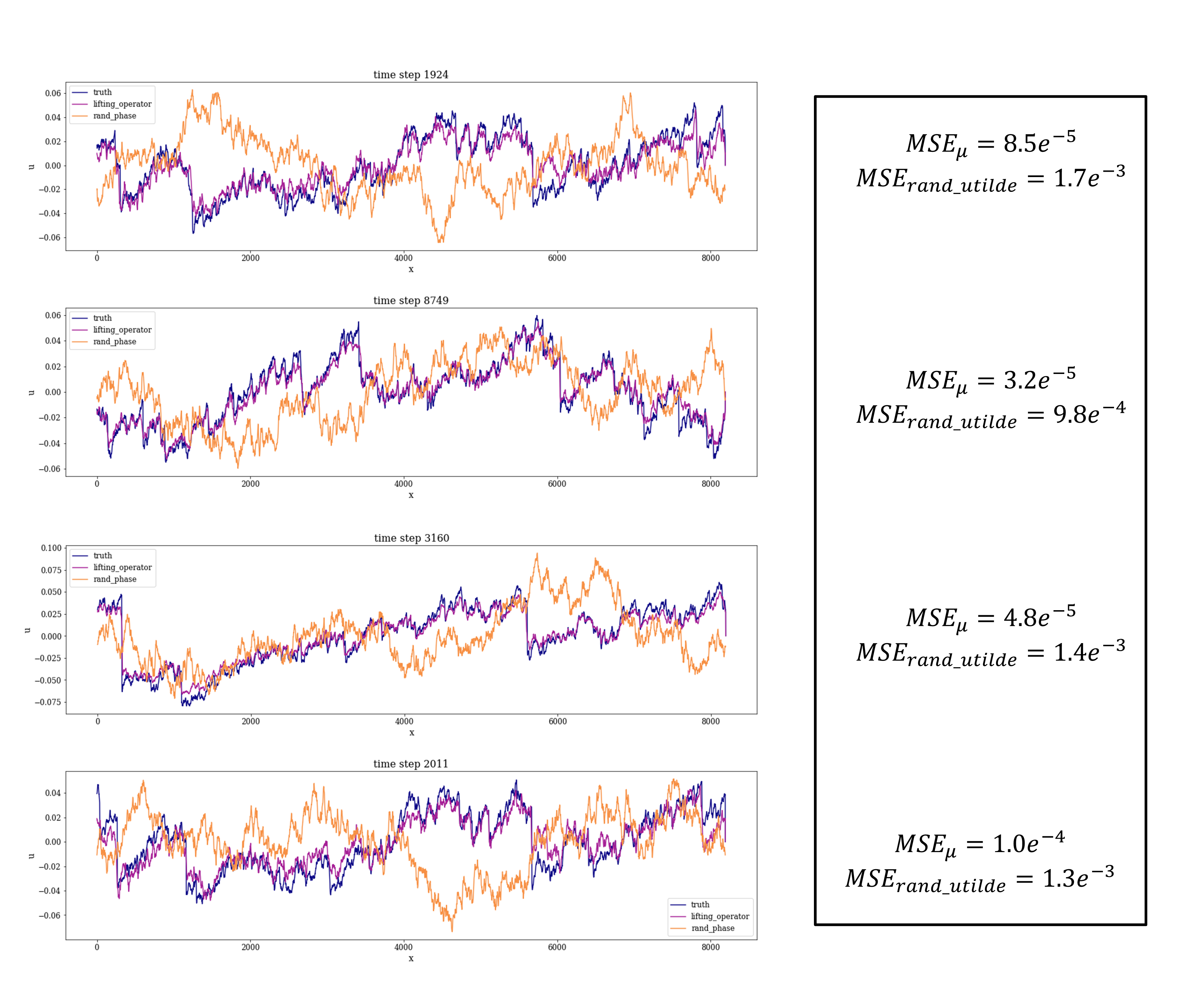}
        \caption{Predictions from the lifting operator at random time steps for an unseen dataset with $\beta$ scaling of 0.99 in the SBE. The figure also compares the accuracy of velocity construction of the lifting operator with the ground truth signal and a random phase initialized velocity signal.}
        \label{fig:lift_op}
    \end{figure}

    The performance of the lifting operator ($\mu$) was evaluated on an unseen dataset generated from the SBE using a different noise scaling value ($\beta$) of 0.99. \textbf{Fig. \ref{fig:lift_op}} demonstrates the accuracy of the velocity signals constructed by the lifting operator at randomly sampled time steps. The lifting operator's predictions were compared to both the ground truth velocity signals from DNS and to velocity signals generated using a randomly-initialized phase. The random-phase velocity signals were generated via the following steps
    
    \begin{gather}
        N_{p} = \frac{N-1}{2} \\
        \theta_{k} = 2\pi \cdot \text{rand()} \quad \text{for } k = 1,\dots,N_{p} \\
        g[k] = f[k] \cdot \left(\cos(\theta_{k}) + j\sin(\theta_{k})\right) \quad \text{for } k = 1,\dots,N_{p} \\
        \tilde{u}_{rand} = \mathcal{F}^{-1}(g),
    \end{gather}
    \noindent where $N_{p}$ is the number of complex elements in the positive half of the frequency domain and $N$ is the complete length of the signal. $f[k]$ represents the amplitude spectrum of the signal at a given time step and $g[k]$ is an array of a positive half complex of numbers calculated by using random phases between 0 and $2\pi$ and by taking $\sin$ and $\cos$ of the phases and combining them as real and imaginary parts, respectively. 'rand()' is a function that generates random values between 0 and 1. Finally, the random signal is calculated by taking the inverse Fourier transform of $g[k]$. In \textbf{Fig. \ref{fig:lift_op}}, four distinct time steps were analyzed to assess the accuracy of the lifting operator's velocity signal approximations. For each time step, the mean squared error (MSE) between the ground truth and the predicted signals was computed for both the lifting operator and the random phase-generated velocity signals. As depicted in the figure, the lifting operator consistently provided more accurate velocity signal approximations compared to the random phase method, as evidenced by the lower MSE values. These results demonstrate the lifting operator's ability to effectively reconstruct velocity signals from the energy spectrum values at different time steps, underscoring its potential utility for the analysis of turbulent flow data.

    \FloatBarrier

\begin{figure}[!h]
      \centering
      \begin{subfigure}[b]{1\textwidth}
        \includegraphics[width=1\textwidth]{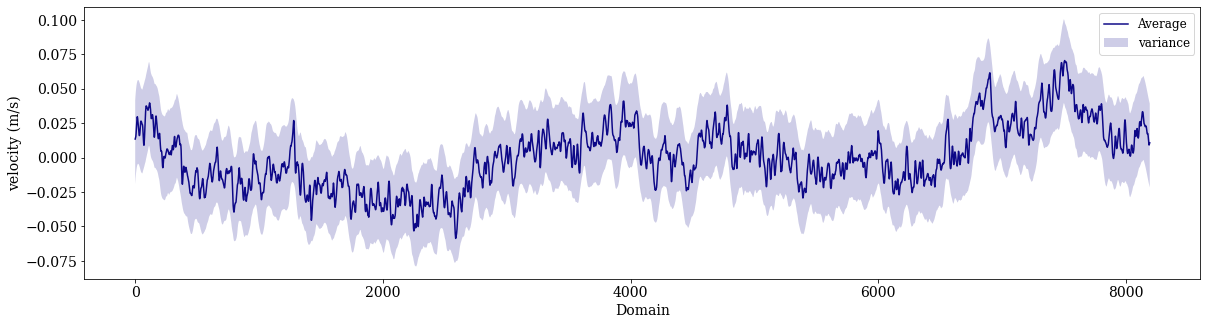}
        \caption{Average and variance of the reconstructed velocity field at $t_{final}$ for five otherwise identical simulations with different initialization of the random initial velocity field.}
        \label{fig:vel}
      \end{subfigure}
      \hfill
      \begin{subfigure}[b]{1\textwidth}
        \includegraphics[width=1\textwidth]{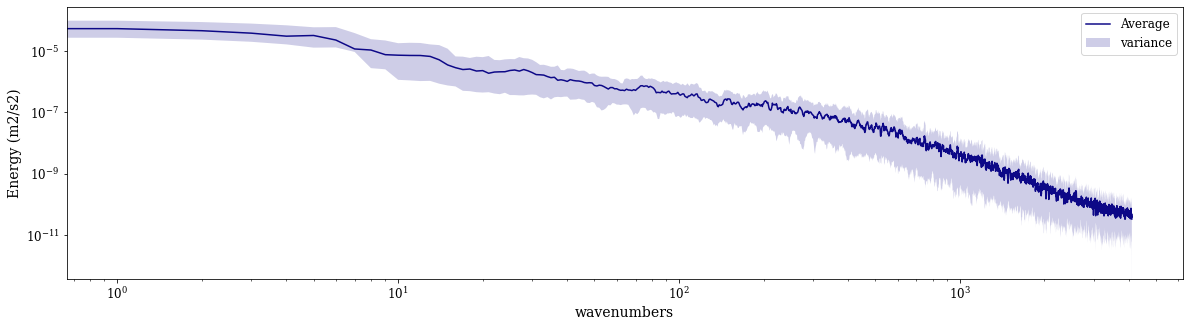}
        \caption{Average and variance of reconstructed energy spectrum signal at $t_{final}$ for five otherwise identical simulations with different initialization of the random initial velocity field.}
        \label{fig:enr}
      \end{subfigure}
      \caption{The behavior of the lifting operator ($\mu$) with different random initializations. The average and variance are calculated for five otherwise identical simulations.}
      \label{fig:figures}
    \end{figure}

    \textbf{Fig. \ref{fig:vel}} and \textbf{Fig. \ref{fig:enr}} show how different random seed initializations of the initial velocity field can affect the accuracy of the velocity field generated by the lifting operator. For a chosen projection ratio ($\frac{dT}{dt}$) of 50 and a healing simulation time of $1e^{5}$ time steps ($t_{heal}$), the generated velocity signal at the final time step will be slightly different for different initializations. To understand to what extent the Burgers' velocity field at the final time step varies with the random initialization, we calculated the nodal velocity variance and mean for ten runs with different random initializations. The restriction operator was then used to calculate the energy spectrum for all the runs. The average and variance were also calculated for the multiple energy spectrum signals.

    \FloatBarrier
    
\begin{figure}[!h]
        \centering
        \includegraphics[width=1\textwidth]{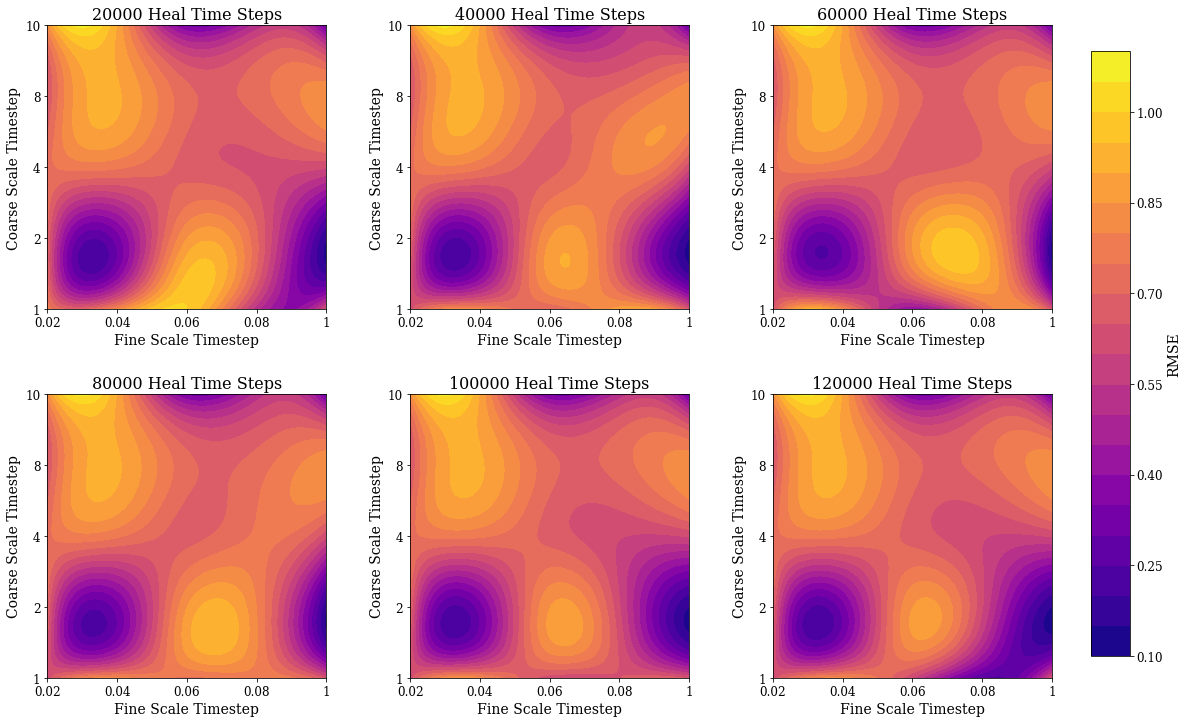}
        \caption{Relative mean square errors (RMSE) between the CPI predicted and DNS only velocity field information at $t_{final}$ for different combinations of $\frac{dT}{dt}$ and $t_{heal}$ times.}
        \label{fig:cp}
    \end{figure}

    To obtain the least discrepancy between the predicted and actual $t_{final}$ velocity signal, the time step ratio and healing times should be chosen carefully. \textbf{Fig. \ref{fig:cp}} shows a comparison between an array of time step ratios and simulation healing times in terms of the overall relative mean squared errors (MSE), which are calculated as follows
    \begin{gather}
        MSE = \frac{1}{n}\sum_{1}^{n}\left(\tilde{u}_{t_{final}}(i) - u_{t_{final}}(i)\right)^{2},    
    \end{gather}
    \noindent where $n$ is the vector length or domain size, $\tilde{u}_{t_{final}}$ is the predicted at $t_{final}$ and healed velocity signal and $u_{t_{final}}$ is the actual DNS velocity signal at $t_{final}$. All of the errors are normalized by the largest MSE in order to find relative MSE. The contour plots presented in \textbf{Fig. \ref{fig:cp}} demonstrate that the regions of low RMSE remain consistent throughout all the healing steps. Notably, as the number of time steps per Coarse Integration Projection (CPI) cycle increases, the areas with relatively high errors are reduced to lower error levels. This observation is evident when comparing the subplot representing 20,000 time steps with the subplot for 120,000 healing time steps. The improvement in error reduction highlights the effectiveness of the healing process in refining the accuracy of the model.

    \FloatBarrier
    
    The contour plots in \textbf{Fig. \ref{fig:cp}} demonstrate a nonlinear relationship between the error in the final velocity field and the coarse-scale and fine-scale time steps. The nonlinear nature of the trend may be attributed to the underlying dynamics of the turbulent system, which often exhibit sensitivity to initial conditions, as well as to intricate interactions between various spatial and temporal scales. Similarly, the complex interplay between the coarse projection step size and healing times reflects the delicate balance between the need for efficient computational performance and the desire for accurate approximations. As the coarse projection step size and healing times are adjusted, the system's behavior may be influenced by the interactions between large-scale energy transfers, small-scale dissipative processes, and the evolving energy spectrum's shape. Furthermore, the nonlinear trend may be indicative of the presence of optimal regions in the parameter space, where the combination of coarse projection step size and healing times minimize the RMSE. Identifying such regions can aid in determining the most appropriate values for these parameters, leading to more accurate and efficient simulations of Burgers turbulence.   
    
    \FloatBarrier

\begin{figure}[!h]
    \centering
    \includegraphics[width=1\textwidth]{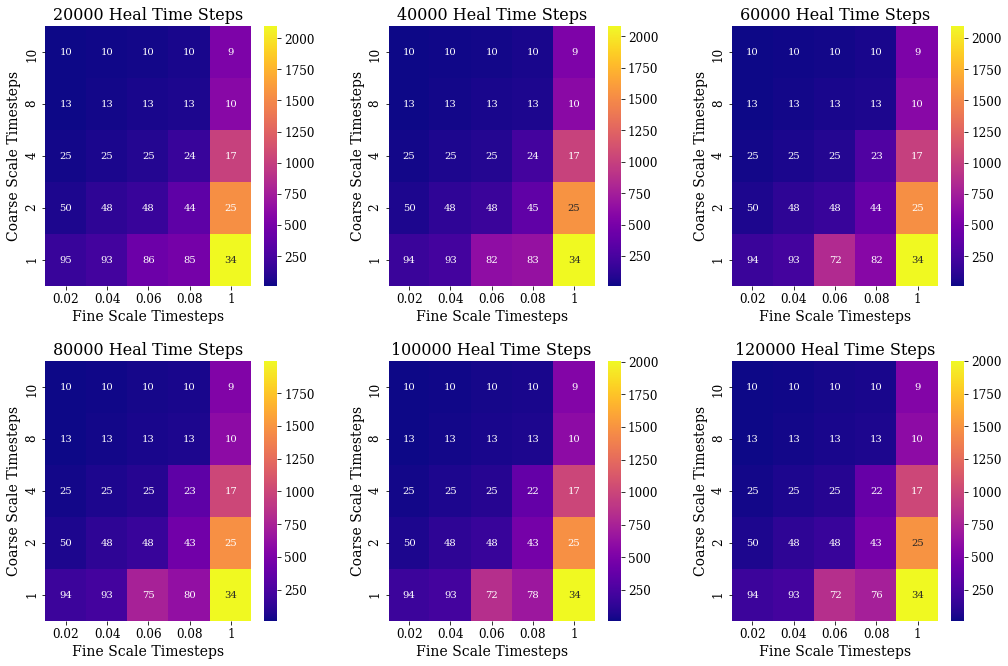}
    \caption{Wall clock time taken to reach $t_{final}$ for different $t_{heal}$ time steps and CPI parameters.}
    \label{fig:euler_time}
\end{figure}

The wall clock time taken by a simulation is dependent upon the coarse projective time steps ($dT$) and simulation healing times ($t_{heal}$) used in the CPI cycles. The relationship is defined as

\begin{gather}
    \log(WCT) = \lambda t_{heal} + \log(\frac{1}{dT}) \label{wct}
\end{gather}
\noindent where $\lambda$ is a constant ranging from 0.63 to 0.72 (based on the simulations performed). As the size of the time stepping size decreases and the healing time increases, the time taken for a simulation to complete increases. Eqn. (\ref{wct}) represents a logarithmic-linear relationship between the Wall Clock time (seconds) and the Simulation Healing Time (seconds) for different coarse projective time stepping values. \textbf{Fig. \ref{fig:euler_time}} shows how does the increase in $t_{heal}$ time steps affects the overall simulation wall clock time for the same CPI parameters selected in \textbf{Fig. \ref{fig:cp}}. In the referenced figure, the subplots illuminate the non-proportional influence of healing time steps on wall clock time across varying magnitudes of coarse projection steps. For larger coarse projection steps, alterations in healing time steps exert negligible effects on the wall clock time. However, when we consider smaller projection steps, an escalation in healing time steps reduces the number of CPI cycles within each simulation, as signified by the numerical values inside the boxes. The changing background box color denotes a corresponding increase in wall clock time.

The preceding discussion of relative mean squared errors illustrates a trade-off: extending the healing steps does curtail the error, yet this benefit is generally outweighed by the corresponding increase in wall clock time. Thus, to maximize the efficiency of our approach, the model should prioritize projection over healing, thereby minimizing the time required to solve the SBE. This strategy ensures that computational resources are directed toward the most beneficial tasks, promoting an optimal balance of accuracy and computational efficiency.

\FloatBarrier

\begin{figure}[!h]
    \centering
    \includegraphics[width=1\textwidth]{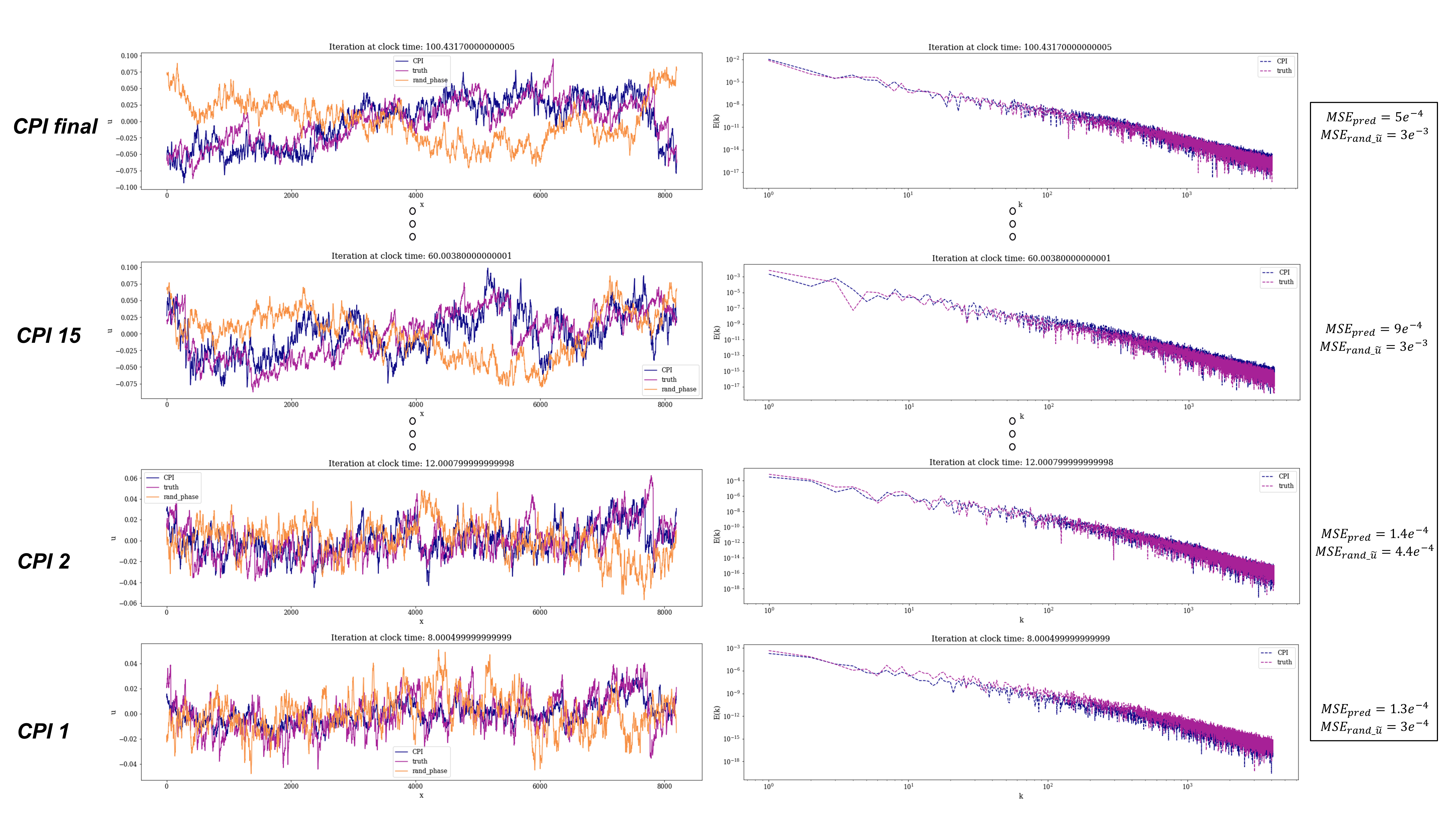}
    \caption{Velocity and energy spectrum reconstruction comparison shown at different CPI steps for a complete simulation. The CPI parameters chosen were: $\frac{dT}{dt} = 2$ and $t_{heal}$ = $1e^{5}$ time steps. The full cycle consisted of 34 CPI cycles.}
    \label{fig:cpi_cycle}
\end{figure}

\textbf{Fig. \ref{fig:cpi_cycle}} shows the coarse projective cycle fine scale information reconstruction for $\frac{dT}{dt} = 2$ and $t_{heal}$ = $1e^{5}$ time steps (10 seconds). The simulation takes about 34 CPI steps (around $1300$ seconds wall clock time) to reach a statistically steady state. The figure also compares the predicted signal with a phase-randomized signal at $t_{final}$ which is healed for the same amount of time as the predicted signal, for comparison. Although the lifting operator demonstrates remarkable performance in approximating the velocity signal given an energy spectrum signal, as depicted in \textbf{Fig. \ref{fig:lift_op}}, errors are introduced through the coarse scale projection. The explicit Euler scheme employed for the projections occasionally leads to inaccuracies, as the projected energy signal may not be close to the true energy spectrum signal. Consequently, the lifting operator's approximations accumulate errors, which can result in longer healing times and a cumulative error addition at each CPI cycle. The impact of these errors on the overall model performance warrants further investigation and potential improvements in the projection methodology.

\FloatBarrier

\begin{figure}[!h]
    \centering
    \includegraphics[width=1\textwidth]{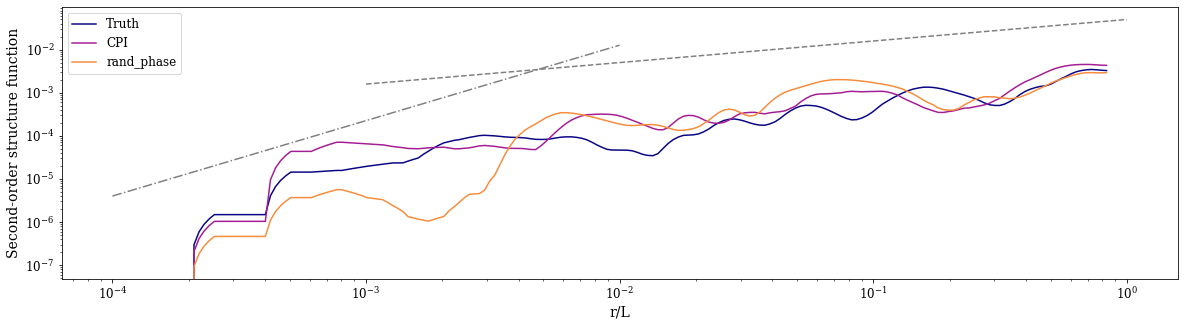}
    \caption{The second-order structure function at $t_{final}$ of the Burgers' velocity field obtain using DNS, the CPI + seq2seq scheme, and a CPI + random phase-initialized velocity scheme, for comparison, which has been averaged over 10 random initialization. The CPI parameters used for the latter two cases were$\frac{dT}{dt} = 2$ and $t_{heal}$ = $1e^{5}$ steps.}
    \label{fig:struct_fn}
\end{figure}

In the figure \textbf{Fig. \ref{fig:struct_fn}}, a comparison between the second-order structural function has been made between the spatial correlations of the reconstructed flow field and the DNS-only flow field. The equation to calculate the second-order structural function is given as

\begin{gather}
    S_2 = \left \langle [u(x+r) - u(x)]^2 \right \rangle    
\end{gather}
\noindent where $u$ is the velocity vector at position $x$, and the angle brackets denote an average over all possible pairs of points separated by a distance $r$. Over the entire flow field, $1000$ nodes were chosen for representing the structural function. The $MSE$ between the ground truth and CPI reconstructed signal is $6e^{-4}$ whereas the $MSE$ between the ground truth and the random-phase initialized signal is $3e^{-3}$.

\FloatBarrier

\section{Summary} \label{summ_sec}

In this work, we make use of the multiscale nature of computationally evolving turbulent flows to accelerate the simulation of the one-dimensional stochastic Burgers' equation to stationarity. We leverage the fact that turbulent flows can be characterized by two distinct spatial and temporal scales, resulting in complex and dynamic behavior that cannot be fully appreciated (or characterized) by considering a single scale alone. Establishing methods to recover a microscopic flow description from a more condensed, macroscopic one poses significant challenges. Earlier analytical models and approximation strategies fell short of effectively encapsulating the inherent physics of the flow. But, due to the advent of state-of-the-art deep learning-based encoder-decoder models, such a translation (and its physics) can be learned following a data-driven approach. Our work, which pairs equation free modeling methods with sequence to sequence learning, demonstrates savings in the computational wall clock time which ranges from a factor of 2 to 442. The Burgers' velocity field reconstruction (or translation) made by the deep learning model is approximately three times more accurate than a randomly phased initialized velocity signal at a given time step. Overall, our research demonstrates the potential of leveraging the multiscale nature of turbulent flows to improve the simulation of complex systems and provides a valuable contribution to the field of computational fluid dynamics and scientific machine learning.


\bibliographystyle{unsrt}
\bibliography{my_refs}

\appendix

\begin{appendices}
		\section{Supplemental information} \label{ase:app_one_sect_1}
\begin{figure}[!h]
                \centering
                \includegraphics[width=1\textwidth]{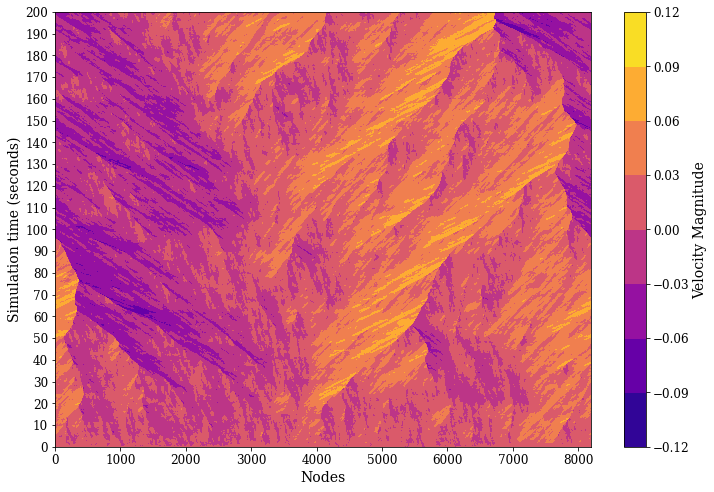}
                \caption{Contour plot showing the evolution of Burgers' velocity field for $\beta = -0.75$.}
                \label{fig:cont_u}
            \end{figure}

            Fig. \ref{fig:cont_u} depicts the evolution of the velocity field for one-dimensional Burgers' turbulence over time. The horizontal axis represents the spatial index, while the vertical axis corresponds to the time index, scaled from 0 to 200. The color intensity in the plot indicates the magnitude of the velocity at each spatial point and at each time step, with a color map ranging from low magnitudes in darker colors to high magnitudes in brighter colors.

            This visualization provides insights into the temporal and spatial dynamics of the velocity field in one-dimensional Burgers' turbulence. We can observe the emergence of distinct features in the velocity field, such as the formation of localized high-velocity regions and the complex interplay between various spatial scales. The contour plot also reveals the non-linear interactions and the self-similarity of the turbulent structures, which are characteristic of Burgers' turbulence.
            
            The analysis of this contour plot helps to understand the underlying mechanisms governing the turbulent flow and can be valuable for further research and development of numerical methods, multiscale modeling, and machine learning techniques to study and simulate turbulence more efficiently.

            \FloatBarrier
            
\begin{figure}[!h]
                \centering
                \includegraphics[width=1\textwidth]{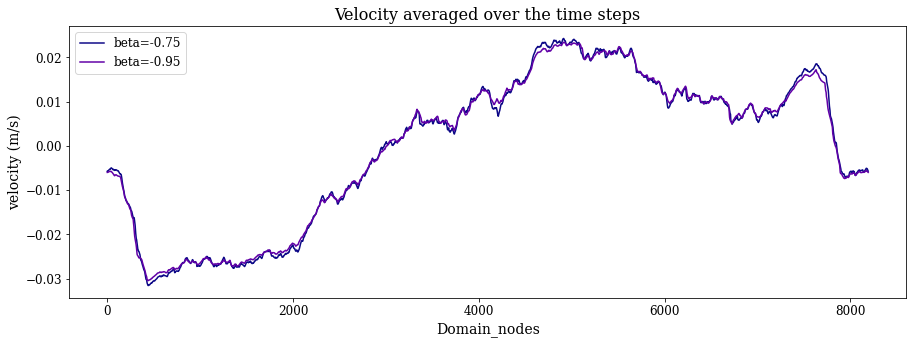}
                \caption{Velocity fields averaged over all time steps for $\beta = -0.75$ and $$-0.95$$.}
                \label{fig:vel_avg}
            \end{figure}

\begin{figure}[!h]
                \centering
                \includegraphics[width=1\textwidth]{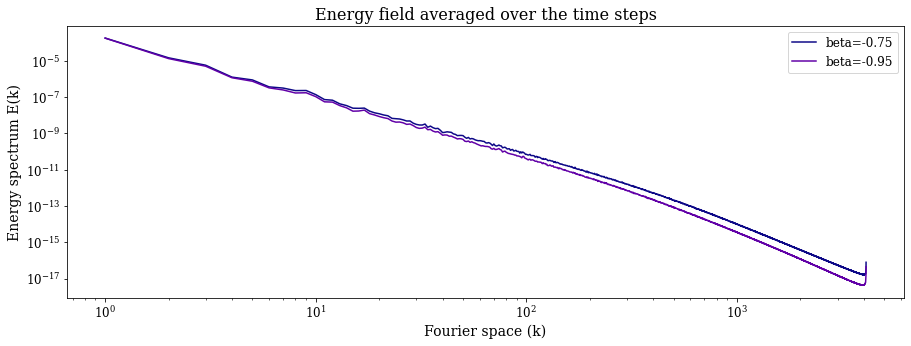}
                \caption{Energy spectra averaged over all time steps for $\beta = -0.75$ and $$-0.95$$.}
                \label{fig:enr_scale}
            \end{figure}

            To understand the general trend over the spatial dimension, Fig. \ref{fig:vel_avg} and Fig. \ref{fig:enr_scale} show an averaged signal distribution over the spatial index and wavenumbers. The dataset generated for the model training was at different values of $\beta$ within the multifractial range.

            \FloatBarrier

\begin{figure}[!h]
                \centering
                \includegraphics[width=1\textwidth]{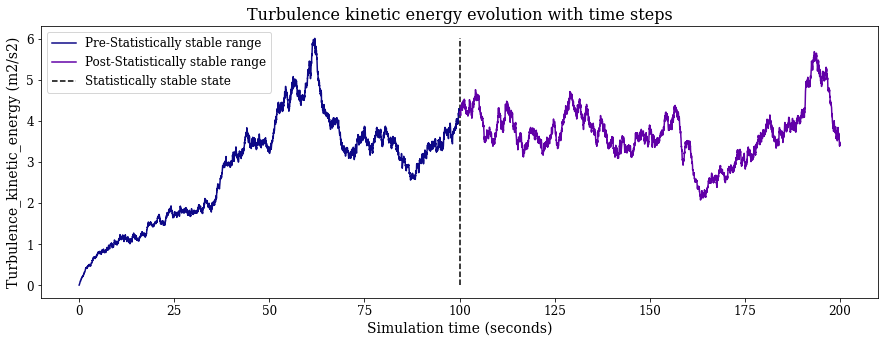}
                \caption{Time evolution of the turbulent kinetic energy (TKE).}
                \label{fig:tke}
            \end{figure}

            The presented plot in Fig. \ref{fig:tke} displays the turbulent kinetic energy (TKE) as a function of time steps for the one-dimensional Burgers' turbulence simulation. The horizontal axis represents the time steps, while the vertical axis corresponds to the TKE values. A clear demarcation line separates the plot into two distinct regions, indicating the point at which the TKE becomes statistically stable.

            In the initial region, the TKE exhibits a transient behavior as the turbulence evolves, characterized by fluctuations and changes in the flow field. This phase represents the early stages of turbulence development, where the flow has not yet reached a statistically steady state. During this period, the TKE experiences variations as the energy cascades through different spatial scales and the flow dynamics gradually settle into a more stable configuration.
            
            Beyond the demarcation line, the TKE enters a statistically stable phase. In this region, the flow has reached a state of statistical equilibrium, and the TKE remains relatively constant. This stability indicates that the energy input from the forcing is balanced by the energy dissipation at the smallest scales of motion.

            By identifying the point at which the TKE becomes statistically stable, we can better understand the turbulence evolution process and focus our efforts on the relevant time periods for studying the underlying physics and improving computational methods for simulating turbulent flows.

            \FloatBarrier

\begin{figure}[!h]
                \centering
                \includegraphics[width=1\textwidth]{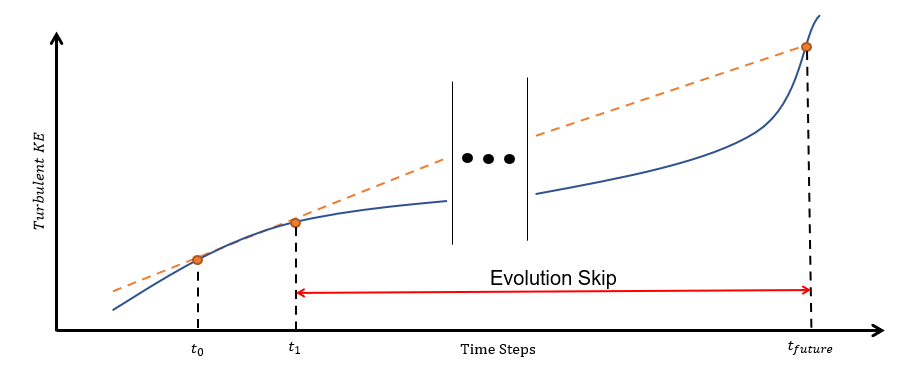}
                \caption{Schematic of the coarse projective integration framework.}
                \label{fig:cpi_sch}
            \end{figure}

            Fig. \ref{fig:cpi_sch} shows a simplified schematic of the coarse projective integration framework.  

            \FloatBarrier

\begin{figure}[!h]
                \centering
                \includegraphics[width=1\textwidth]{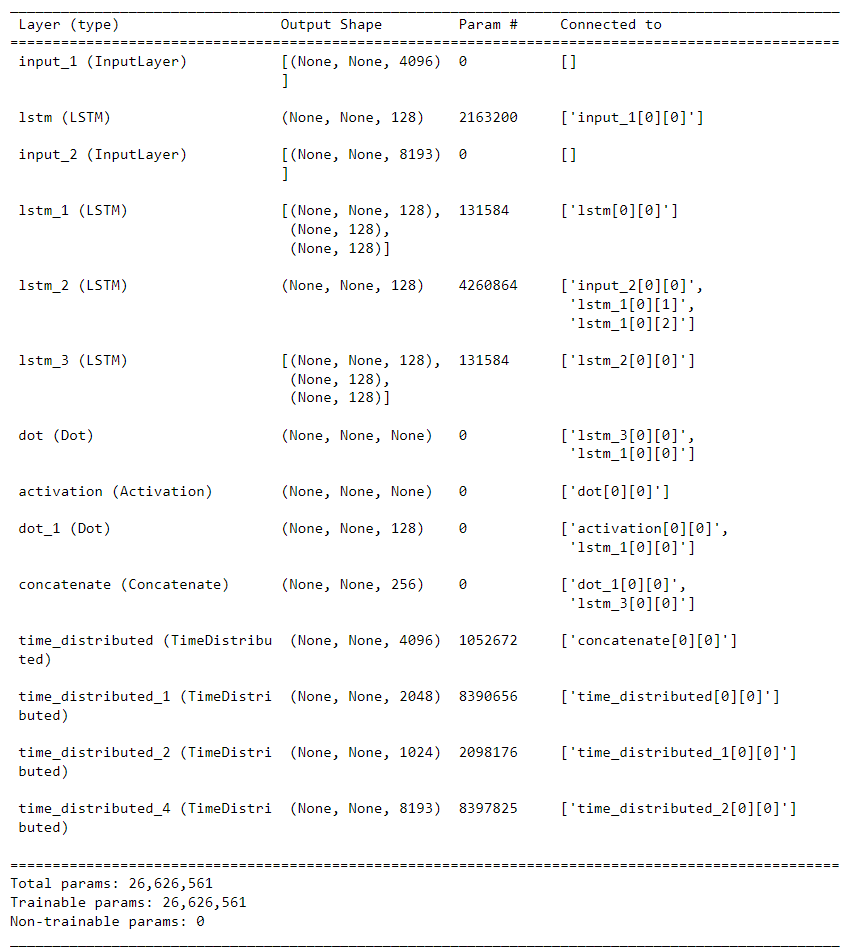}
                \caption{The encoder-decoder model parameters used.}
                \label{fig:model_summary}
            \end{figure}

            The specifications of the encoder-decoder model employed for the sequence-to-sequence learning task in this paper are shown in Fig. \ref{fig:model_summary}. The summary plot presents a comprehensive visual representation of the encoder-decoder model used in this study for simulating one-dimensional Burgers' turbulence. This plot provides valuable insights into the model's architecture, the number of layers, the number of hidden units, and other essential parameters that contribute to the model's performance. 

            \FloatBarrier
            
\begin{figure}[!h]
                \centering
                \includegraphics[width=1\textwidth]{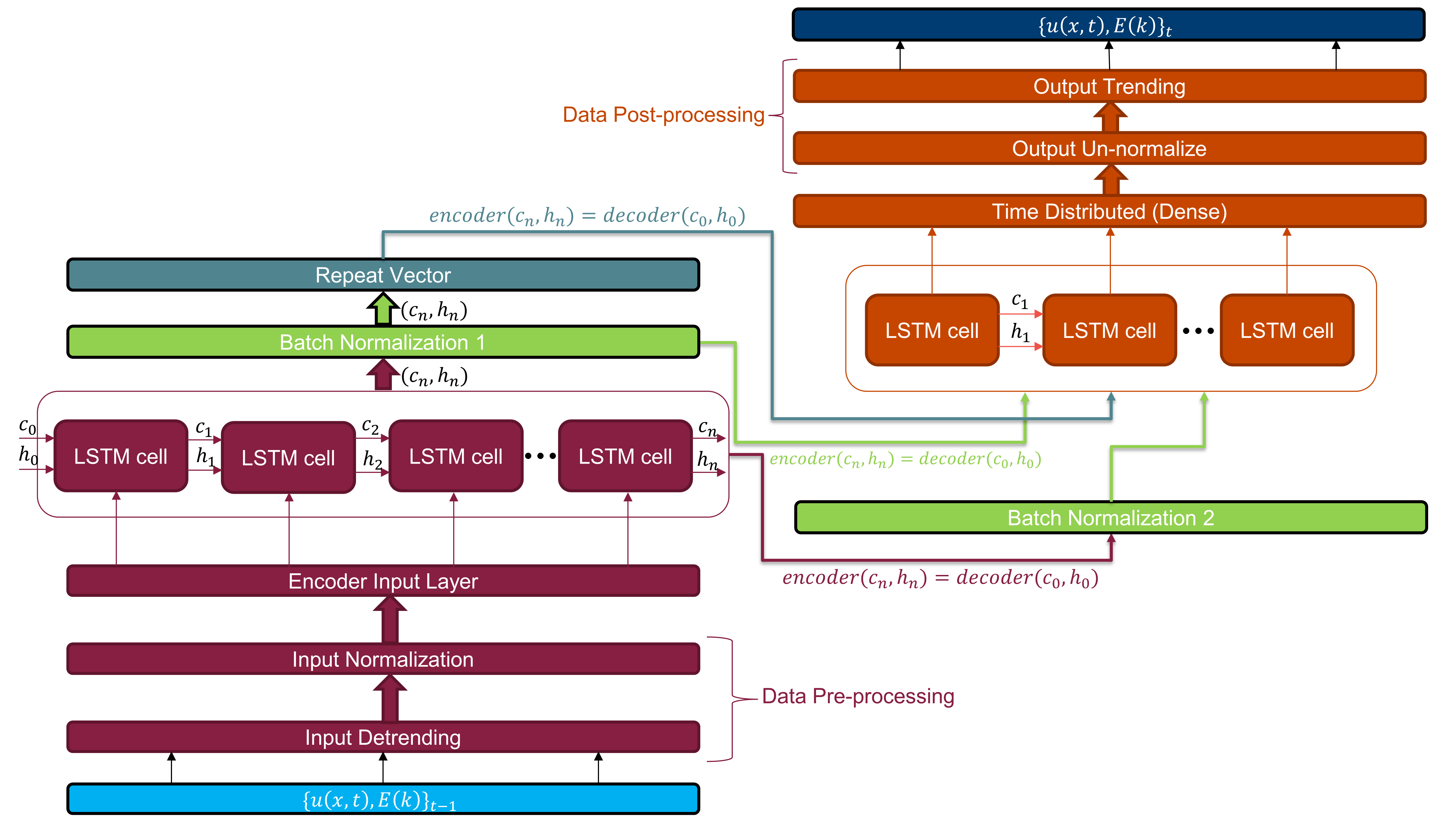}
                \caption{An earlier implementation of the encoder-decoder model.}
                \label{fig:old_model}
            \end{figure}

\begin{figure}[!h]
                \centering
                \includegraphics[width=1\textwidth]{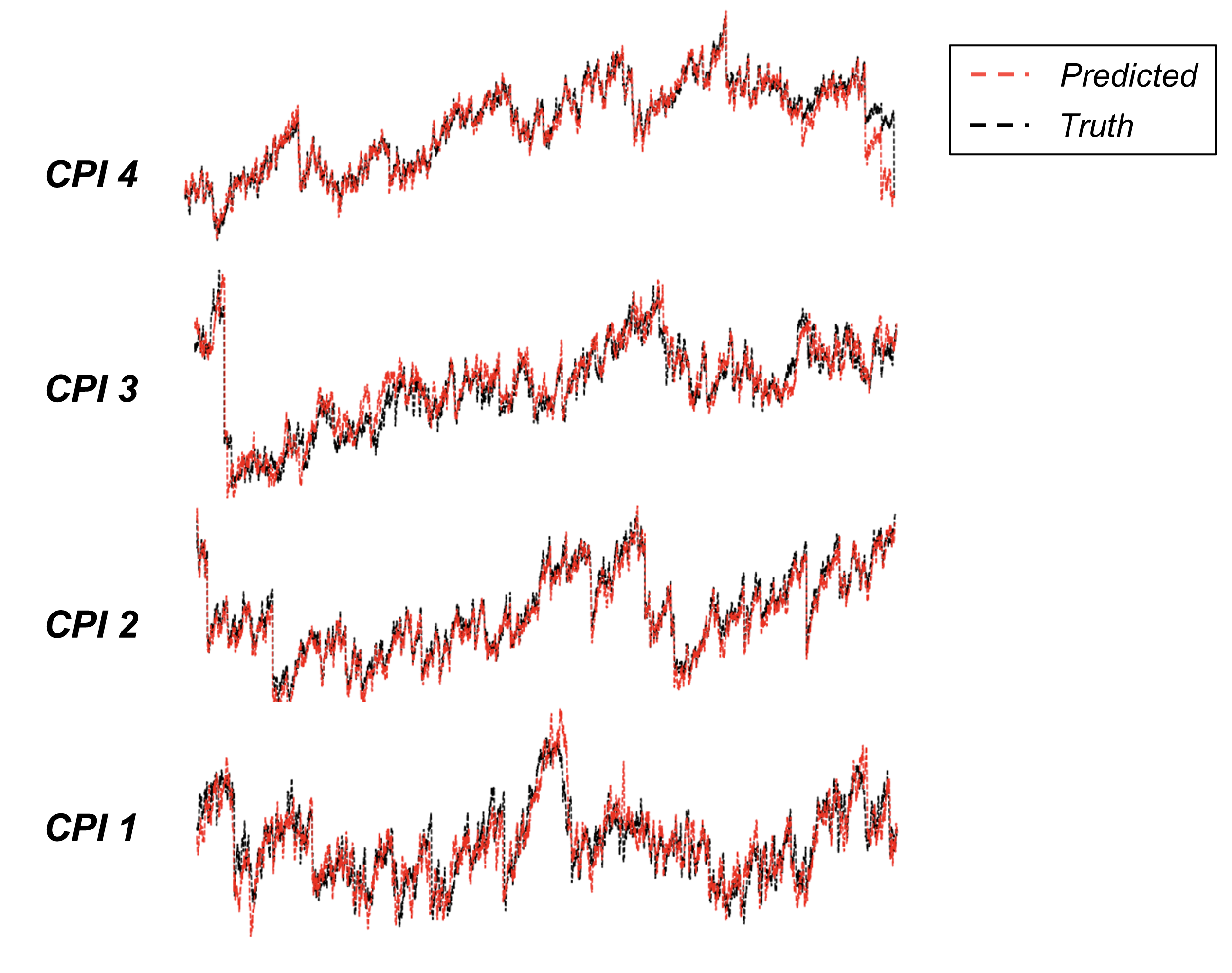}
                \caption{Comparison of the reconstructed velocity signal with the ground truth obtained via DNS for an earlier lifting operator implementation.}
                \label{fig:cpi_old}
            \end{figure}

            The model architecture shown in Fig. \ref{fig:old_model} was one of the early implementations for a lifting operator to transit between the energy spectrum and velocity field. The approach is slightly different from the one demonstrated in the paper as for this model to work, we require a fraction of the energy wavenumber and velocity nodal data points. This fraction is then used to train the model and then the inferences were made of the remaining data points. This approach was not ideal for the CPI scheme as we only have the velocity flow information from a past time step and would like to lift the projected energy spectrum information back to its corresponding velocity field. Fig. \ref{fig:cpi_old} shows the results from this approach and model. Although the $MSE$ between the ground truth and reconstructed signal is really small, this approach was relatively more time-consuming than the one presented in this paper as it requires a model training cycle at each time step.

\end{appendices}

\end{document}